\documentclass[conference]{IEEEtran}
\IEEEoverridecommandlockouts
\usepackage{cite}
\usepackage{amsmath,amssymb,amsfonts}
\usepackage{algorithmic}
\usepackage{graphicx}
\usepackage{textcomp}
\usepackage{xcolor}
\def\BibTeX{{\rm B\kern-.05em{\sc i\kern-.025em b}\kern-.08em
    T\kern-.1667em\lower.7ex\hbox{E}\kern-.125emX}}

\usepackage{listings, xcolor}
\definecolor{verylightgray}{rgb}{.97,.97,.97}
\lstdefinelanguage{Solidity}{
  keywords=[1]{anonymous, assembly, assert, balance, break, call, callcode, case, catch, class, constant, continue, constructor, contract, debugger, default, delegatecall, delete, do, else, emit, event, experimental, export, external, false, finally, for, function, gas, if, implements, import, in, indexed, instanceof, interface, internal, is, length, library, log0, log1, log2, log3, log4, memory, modifier, new, payable, pragma, private, protected, public, pure, push, require, return, returns, revert, self-destruct, send, solidity, storage, struct, suicide, super, switch, then, this, throw, transfer, true, try, typeof, using, value, view, while, with, addmod, ecrecover, keccak256, mulmod, ripemd160, sha256, sha3}, 
  keywordstyle=[1]\color{blue}\bfseries,
  keywords=[2]{address, bool, byte, bytes, bytes1, bytes2, bytes3, bytes4, bytes5, bytes6, bytes7, bytes8, bytes9, bytes10, bytes11, bytes12, bytes13, bytes14, bytes15, bytes16, bytes17, bytes18, bytes19, bytes20, bytes21, bytes22, bytes23, bytes24, bytes25, bytes26, bytes27, bytes28, bytes29, bytes30, bytes31, bytes32, enum, int, int8, int16, int24, int32, int40, int48, int56, int64, int72, int80, int88, int96, int104, int112, int120, int128, int136, int144, int152, int160, int168, int176, int184, int192, int200, int208, int216, int224, int232, int240, int248, int256, mapping, string, uint, uint8, uint16, uint24, uint32, uint40, uint48, uint56, uint64, uint72, uint80, uint88, uint96, uint104, uint112, uint120, uint128, uint136, uint144, uint152, uint160, uint168, uint176, uint184, uint192, uint200, uint208, uint216, uint224, uint232, uint240, uint248, uint256, var, void, ether, finney, szabo, wei, days, hours, minutes, seconds, weeks, years},  
  keywordstyle=[2]\color{teal}\bfseries,
  keywords=[3]{block, blockhash, coinbase, difficulty, gaslimit, number, timestamp, msg, data, gas, sender, sig, value, now, tx, gasprice, origin},  
  keywordstyle=[3]\color{violet}\bfseries,
  identifierstyle=\color{black},
  sensitive=false,
  comment=[l]{//},
  morecomment=[s]{/*}{*/},
  commentstyle=\color{gray}\ttfamily,
  stringstyle=\color{red}\ttfamily,
  morestring=[b]',
  morestring=[b]"
}
\usepackage[hyphens]{url}
\usepackage{breakurl}
\usepackage{graphicx}

\begin{document}

\title{Definition and Detection of Centralization Defects in Smart Contracts}

\author{\IEEEauthorblockN{Zewei Lin}
\IEEEauthorblockA{\textit{School of Software Engineering} \\
\textit{Sun Yat-sen University}\\
Zhuhai, China
}
\IEEEauthorblockA{\textit{Peng Cheng Laboratory} \\
Shenzhen, China \\
linzw3@mail2.sysu.edu.cn}
\and
\IEEEauthorblockN{Jiachi Chen}
\IEEEauthorblockA{\textit{School of Software Engineering} \\
\textit{Sun Yat-sen University}\\
Zhuhai, China \\
chenjch86@mail.sysu.edu.cn}
\and
\IEEEauthorblockN{Jiajing Wu}
\IEEEauthorblockA{\textit{School of Software Engineering} \\
\textit{Sun Yat-sen University}\\
Zhuhai, China \\
wujiajing@mail.sysu.edu.cn}
\and
\IEEEauthorblockN{Weizhe Zhang}
\IEEEauthorblockA{\textit{School of Computer Science and Technology} \\
\textit{Harbin Institute of Technology}\\
Shenzhen, China
}
\IEEEauthorblockA{\textit{Peng Cheng Laboratory} \\
Shenzhen, China \\
wzzhang@hit.edu.cn}
\and
\IEEEauthorblockN{Zibin Zheng*}
\IEEEauthorblockA{\textit{School of Software Engineering} \\
\textit{Sun Yat-sen University}\\
Zhuhai, China \\
zhzibin@mail.sysu.edu.cn}
}

\maketitle

\begin{abstract}
In recent years, security incidents stemming from centralization defects in smart contracts have led to substantial financial losses. A centralization defect refers to any error, flaw, or fault in a smart contract's design or development stage that introduces a single point of failure. Such defects allow a specific account or user to disrupt the normal operations of smart contracts, potentially causing malfunctions or even complete project shutdowns. Despite the significance of this issue, most current smart contract analyses overlook centralization defects, focusing primarily on other types of defects.
To address this gap, our paper introduces six types of centralization defects in smart contracts by manually analyzing 597 Stack Exchange posts and 117 audit reports. For each defect, we provide a detailed description and code examples to illustrate its characteristics and potential impacts. Additionally, we introduce a tool named CDRipper (\textbf{C}entralization \textbf{D}efects \textbf{Ripper}) designed to identify the defined centralization defects.
Specifically, CDRipper constructs a permission dependency graph (PDG) and extracts the permission dependencies of functions from the source code of smart contracts. It then detects the sensitive operations in functions and identifies centralization defects based on predefined patterns.
We conduct a large-scale experiment using CDRipper on 244,424 real-world smart contracts and evaluate the results based on a manually labeled dataset. Our findings reveal that 82,446 contracts contain at least one of the six centralization defects, with our tool achieving an overall precision of 93.7\%.
\end{abstract}


\section{Introduction} \label{sec:intro}
The rapid development of Decentralized Applications (DApps) and Decentralized Finance (DeFi) has spurred extensive research on smart contracts. Recently, a notable increase in security incidents and economic losses has been attributed to centralization defects in DeFi projects. According to the Certik's report~\cite{certikreport2023}, during the third quarter of 2023 alone, more than 14 related incidents were reported, leading to a total loss exceeding \$204 million.

A centralization defect in a DeFi project refers to any error, flaw, or fault in a smart contract's design or development stage that results in a single point of failure~\cite{certikblog2022}. This means that specific accounts, users, or addresses could disrupt normal operations, potentially causing project malfunctions or even shutdown.
These defects encompass centralization issues and logical errors that directly result in the loss of funds. For instance, if the contract owner has the privilege to transfer tokens deposited by users, all users' assets may be at risk as the private key of the owner could be leaked. Additionally, centralization defects can also stem from design flaws that may not pose threats to users' funds immediately but do affect safety in certain scenarios.
In March 2024, the DeFi project \textit{MUMI} incurred an economic loss due to a centralization defect. The contract included a function for minting \textit{MUMI} tokens, which was controlled by a single node. Minting tokens involves generating new tokens by authenticating data, creating new blocks, and recording this information on the blockchain~\cite{mint_def}. An attacker exploited this function to clandestinely mint and drain a substantial number of tokens, resulting in a financial loss of approximately \$35,000~\cite{moti_example}.

Although previous works~\cite{chen2020defining,yang2023definition} have reported a set of smart contract defects, the unique challenges presented by centralization defects have not been fully addressed. To address this gap, we first conducted an empirical study aimed at defining centralization defects in smart contracts. Our analysis encompassed 597 Q\&A posts collected from Ethereum Stack Exchange~\cite{etherstack} and 117 smart contract audit reports from blockchain security companies. By employing a manual filtering and open card sorting approach, we defined six centralization defects in smart contracts: \textit{Mint Function with Single Signature}, \textit{Management without Timelock}, \textit{Critical Variables Manipulation with Single Signature}, \textit{Single Proxy Admin}, \textit{Self-destruct with Single Signature}, and \textit{Individual Contract Output Reliance}. 

To identify these centralization defects, we developed a tool named CDRipper (\textbf{C}entralization \textbf{D}efects \textbf{Ripper}), which takes the smart contract's source code as input.
By identifying centralization defects in contracts, developers can enhance contract security while users can mitigate investment risks, contributing to the safe and sustainable growth of the smart contract community.
CDRipper first constructs a permission dependency graph (PDG), which represents the relationships of permission dependencies among contract statements and functions, such as those used for or controlled by multi-signature verification, timelock mechanism, and permission checks.
CDRipper then identifies sensitive operations in functions based on predefined rules, which are summarized from the empirical study of centralization defects.
Finally, CDRipper reports centralization defects based on the permission dependencies of functions and sensitive operations.

We applied CDRipper to 244,424 real-world smart contracts, discovering that 82,446 (about 33.73\%) contracts in our dataset contained at least one centralization defect.
More than 80\% of reported defects arise from management through a single-signature address, i.e., an address controlled by a single private key. 
The compromise or loss of the private key associated with this address will leave the contract open to unauthorized access and lead to security issues. In the performance evaluation of CDRipper, we used a random sampling method and manually labeled two distinct datasets. The evaluation results show that CDRipper achieved an overall precision rate of 93.7\% and a false negative rate of 14.6\%.

The main contributions of our work are as follows:
\begin{itemize}
\item We defined six types of centralization defects in smart contracts, which is the most comprehensive work on this topic. For each defect, we present an illustration and a code example. Additionally, we provide possible solutions aimed at enhancing the security of smart contracts.
\item We developed a tool called CDRipper for detecting the defined centralization defects. Through random sampling and manual validation, the results demonstrate that our method achieves an overall precision of 93.7\% and a false-negative rate of 14.6\%. 
\item We conducted a large-scale experiment involving 244,424 real-world smart contracts to evaluate the performance of CDRipper. Our findings reveal that among these smart contracts, 82,446 (about 33.73\%) contain at least one defined defect.
\item We published the source code of CDRipper, along with all experimental data and analysis results at: \textbf{https://anonymous.4open.science/r/CDRipper/}.
\end{itemize}

\section{Background} \label{sec:bkg}
To facilitate a better understanding of centralization defects, we provide essential background information in this section.
\subsection{Smart Contracts}
A smart contract serves as a computerized transaction protocol that autonomously enforces the terms of a contractual agreement~\cite{szabo1997formalizing}. 
Ethereum stands out as one of the most widely adopted platforms for smart contracts~\cite{zheng2020overview}.
When a smart contract is deployed to Ethereum, its source code undergoes compilation by the Ethereum Virtual Machine (EVM) compiler~\cite{evm}. This compilation process yields bytecode, which is subsequently stored on the blockchain. Variables declared within a contract but outside any function are denoted as state variables~\cite{statvar}. Notably, state variables are permanently stored in the EVM's storage, rather than in its memory.



\subsection{Decentralized Ecosystem}
Unlike a centralized ecosystem, a decentralized ecosystem functions independently of specific centralized nodes. Governance within the decentralized ecosystem is distributed among its creators and users. Due to the characteristics inherent in a decentralized ecosystem, such as verifiability, self-governance, permissionlessness, and native payments, it enables universal access to services without the requirement for personal data.

\begin{figure}[h]
    \centering
    \setlength{\abovecaptionskip}{0.05cm}
    \includegraphics[width=0.85\linewidth]{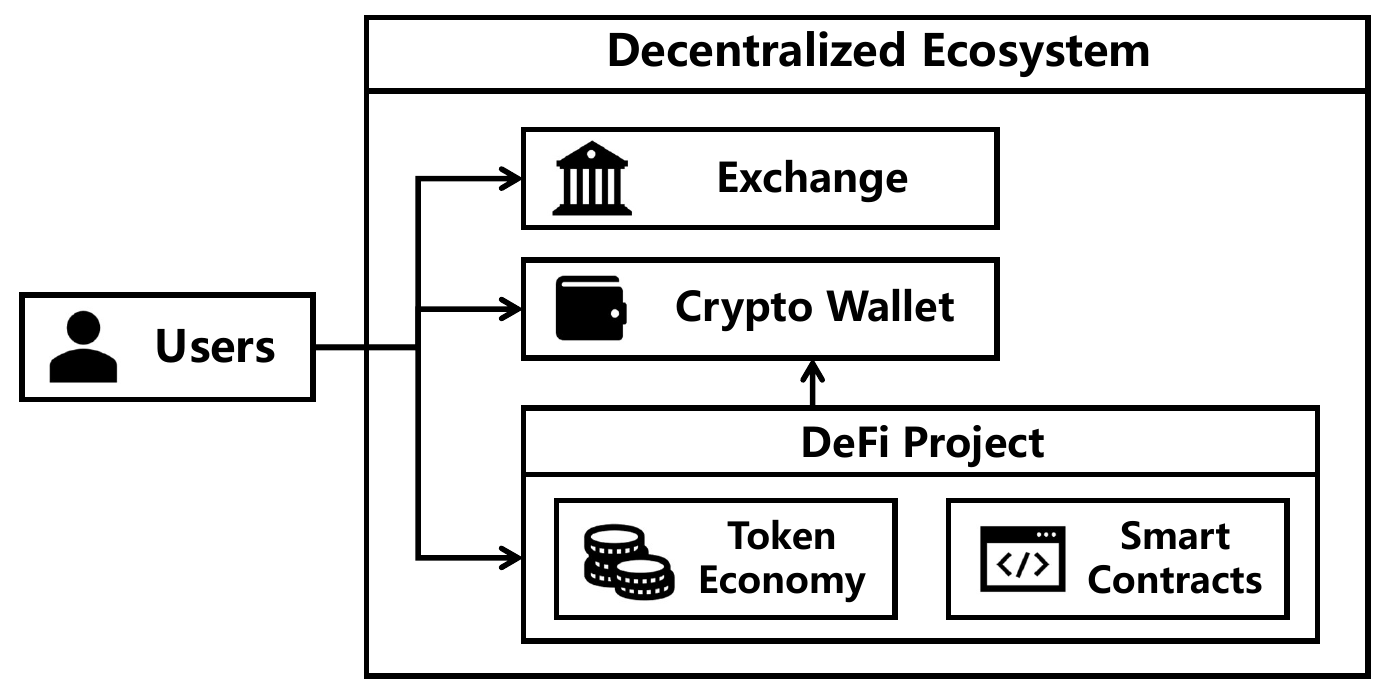}
    \caption{Basic components of a decentralized ecosystem}
    \label{fig:bkg_decen}
\end{figure}

Figure~\ref{fig:bkg_decen} depicts the primary components of a decentralized ecosystem~\cite{yan2023bad}, including Exchanges, which facilitate the trading of fiat and cryptocurrencies; crypto wallets, which serve to manage crypto assets and interact with DeFi projects. 
The arrows in the figure represent the invocation relationships between the components.
The DeFi Project comprises a token economy and smart contracts. The token economy delineates the distribution strategy of the DeFi project's tokens, specifying the allocated percentage for providing liquidity in exchanges and other purposes. Smart contracts are employed to automatically execute the logic of the DeFi project. 

\vspace{-0.3cm}
\vspace{-0.3cm}

\subsection{Multi-signature Wallet}
The private key is an alphanumeric code in the field of cryptography~\cite{privarkey}. Similar to a password, it plays a crucial role in securing and validating cryptocurrency transactions by providing evidence of ownership and facilitating cryptographic authorization~\cite{multisig}.

A single-signature address relies on a single private key. Multi-signature wallets require a specific number of signers to collectively sign a transaction using their respective private keys for approval. The process of verifying whether the number of signers meets the predefined threshold is called multi-signature verification.

\section{Centralization Defects in Smart Contracts} \label{sec:defects}

In response to the insufficient research on centralization defects, we conducted an empirical study and classified them into six types.
In this section, we will illustrate the process of identifying the six centralization defects, providing a definition and example for each one.

\subsection{Data Collection} \label{sec:defect_collect}
\subsubsection{Ethereum Stack Exchange Posts}
To identify centralization defects in smart contracts, it is essential to gain insights into the centralization concerns acknowledged by smart contract developers. Ethereum Stack Exchange~\cite{etherstack} serves as a question-and-answer platform for Ethereum users. 
We conducted searches using the keywords ``centralization'', ``decentralization'', and ``smart contract'' up to October 2023.
Ultimately, we collected 597 posts pertaining to defects related to centralization in smart contracts.

\subsubsection{Smart Contract Audit Reports}
In addition to posts, smart contract audit reports also serve as a valuable resource for identifying centralization defects in smart contracts. In recent years, with the increasing economic impact of centralization defects in smart contracts, certain smart contract security platforms have incorporated the assessment of centralization risk as part of their audit criteria. 
We collected audit reports from various smart contract security auditing platforms up to October 2023, including Certik~\cite{certik}, SourceHat~\cite{sourcehat}, and others.
Subsequently, we filtered out reports that had audited centralization defects. Ultimately, we collected 117 audit reports addressing centralization defects.

\subsection{Data Analysis} \label{sec:defect_analysis}
\subsubsection{Manually Filtering} 
Not all of the posts and audit reports are relevant to smart contract centralization defects.
For instance, some posts delve into the centralization defects of cross-chain protocols or off-chain wallets. The root cause of these defects lies in the design of the protocol rather than in the development of smart contracts, placing them beyond the scope of this paper. Furthermore, audit reports may not identify any centralization defects in the audited smart contracts. Therefore, we manually excluded data unrelated to smart contract centralization defects. Finally, we identified that 139 posts and 49 audit reports were relevant to Solidity-related smart contract centralization defects.

\subsubsection{Open Card Sorting}
To ensure accuracy, we employed the open card sorting method~\cite{spencer2009card} to analyze and categorize the filtered posts and audit reports. A card was generated for each post or report, with its content divided into several parts for convenient analysis, i.e., title, description, comments, or recommendations. Two researchers, each possessing over two years of experience in smart contract research, collaborated on the analysis and classification. We conducted a total of two rounds of classification as follows:

In the first round, we randomly selected 40\% of the cards, and the two researchers collaborated to analyze and determine the classification for these cards. In the first step, they read the title and description of each card to understand the defects associated with it. Subsequently, they reviewed the comments or recommendations to understand how to address the identified defects. Furthermore, cards without a clear root cause of defects were omitted.

In the second round, the two researchers independently analyzed and categorized the remaining 60\% of the cards. The detailed analysis steps were the same as in the first round. Subsequently, they compared the results and discussed the differences, either harmonizing or deleting them after the discussion. Finally, they classified the smart contract centralization defects into six types.

\vspace{-0.4cm} 
\begin{figure}[h]
    \centering
    \setlength{\abovecaptionskip}{0.05cm}
    \includegraphics[width=\linewidth]{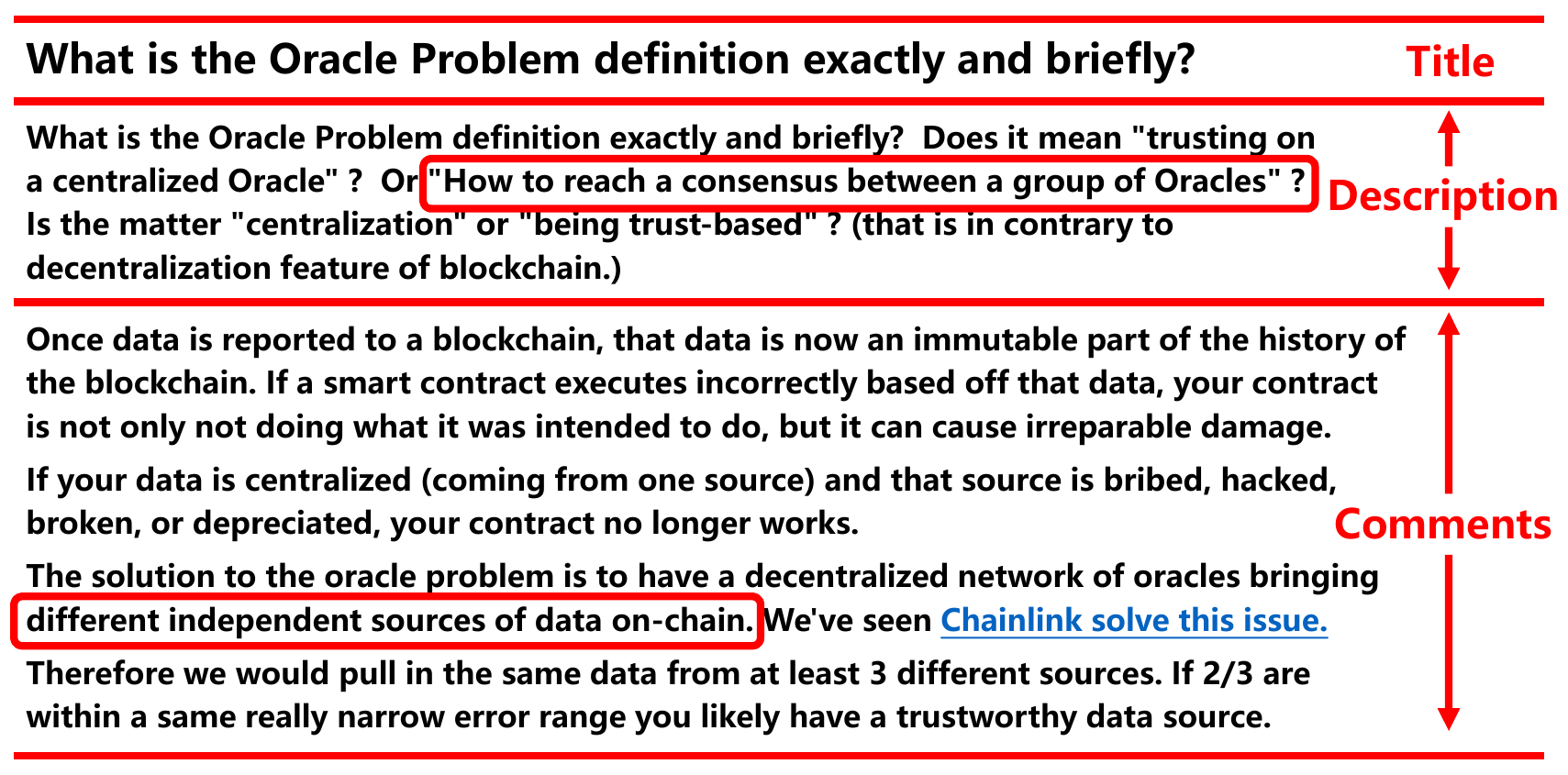}
    \caption{Example of a card of Ethereum Stack Exchange posts}
    \label{fig:defect_posts}
\end{figure}
\vspace{-0.3cm} 

Figure \ref{fig:defect_posts} illustrates a card example generated from an Ethereum Stack Exchange post~\cite{post_oracle}.
The card comprises three parts: the title, description, and comments.
In the description, the questioner raised doubts about the definition of oracle problems and reaching a consensus among a group of oracles. In response, the comments emphasized the risk of relying on data from a single source and recommended the use of various independent data sources. Consequently, we classified this issue as ``Individual Contract Output Reliance''

\begin{figure}[h]
    \centering
    \setlength{\abovecaptionskip}{0.05cm}
    \includegraphics[width=\linewidth]{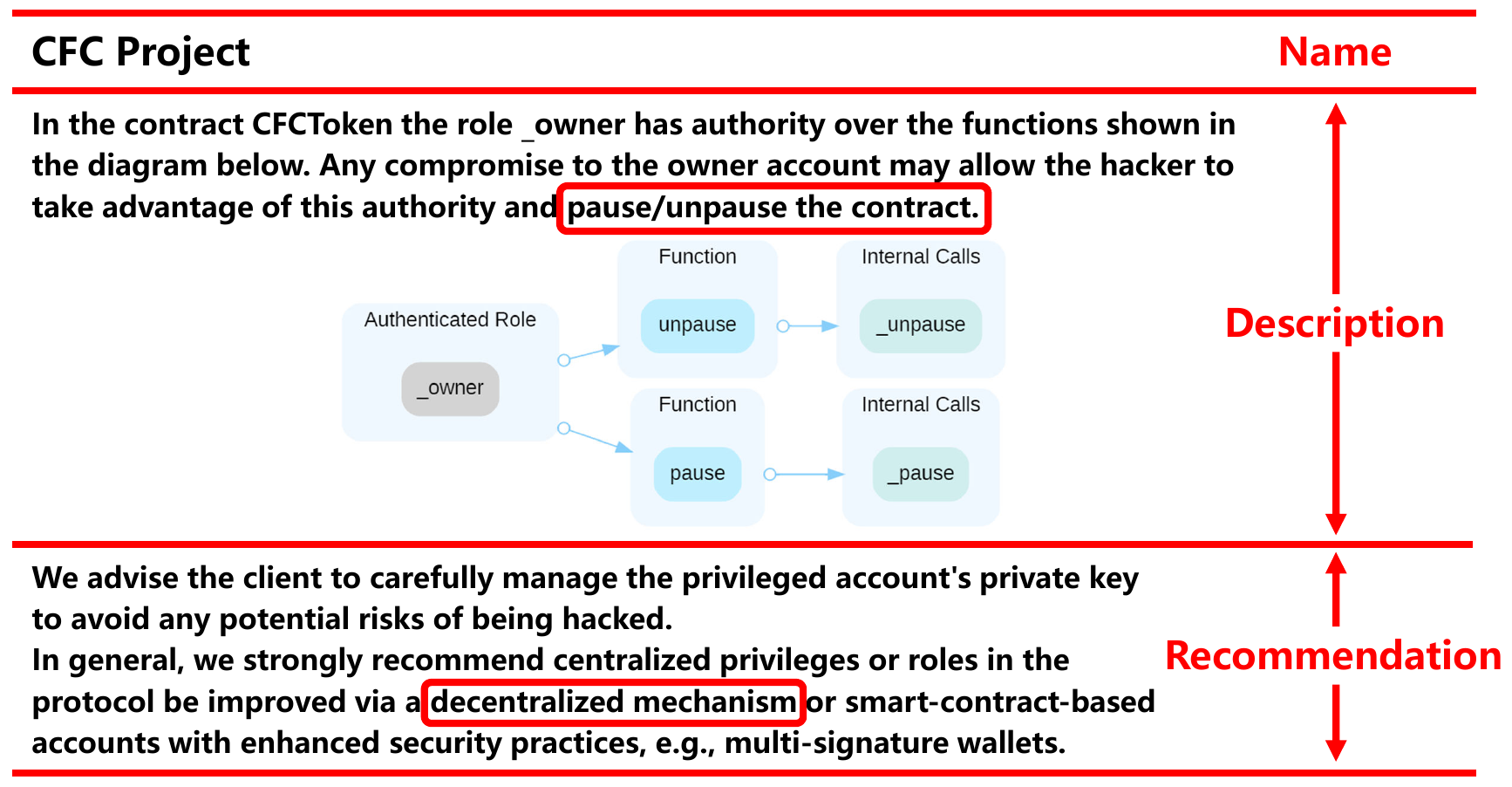}
    \caption{Example of a card of smart contract audit reports}
    \label{fig:defect_report}
\end{figure}

Figure \ref{fig:defect_report} illustrates an example of a card generated from a smart contract audit report~\cite{report_cfc}. 
The card comprises three parts: the project name, description, and recommendation. As indicated in the description, there are functions that allow the modification of critical variables in the smart contract, and these functions can only be invoked by the role \textit{\_owner}. Compromising the \textit{\_owner} account may enable a hacker to exploit this authority and pause or unpause the contract. The audit platform recommends mitigating this risk by employing a decentralized mechanism, such as multi-signature wallets or timelock smart contracts. Consequently, we categorize this issue as ``Critical Variables Manipulation with Single Signature''.

\subsection{Centralization Defects Definition}
\label{sec:definition}
Based on open card sorting, we finally defined six types of smart contract centralization defects. 
We provide a brief definition of these smart contract centralization defects in Table \ref{tab:defects_def}. Subsequently, we will elaborate on the detailed definitions and present code examples of these defects later in this subsection.

\vspace{-0.4cm}
\begin{table}[h] 
    \centering
    \setlength{\abovecaptionskip}{0.05cm}
        \caption{Definition of centralization defects.}
        \resizebox{\linewidth}{!}{
        \begin{tabular}{p{3.6cm}|p{5.2cm}}
            \hline
            \textbf{Centralization Defect} & \textbf{Definition} \\
            \hline
            \textit{Mint Function with Single Signature (MFS)} & There are functions for arbitrary token minting controlled by a single-signature address. \\
            \hline
            \textit{Critical Variables Manipulation with Single Signature (CVS)} & Critical Variables can be manipulated with a single-signature address.\\
            \hline
            \textit{Management without Timelock (MT)} & Management functions can be executed without time delay. \\
            \hline
            \textit{Single Proxy Admin (SPA)} & The admin of the proxy contract is a single-signature address. \\
            \hline
            \textit{Self-destruct with Single Signature (SS)} & The self-destruct function can be executed by a single-signature address. \\
            \hline
            \textit{Individual Contract Output Reliance (IOR)} & Existing logic relies on the output of an individual external contract. \\
            \hline
        \end{tabular}}
    \label{tab:defects_def}
\end{table}
\vspace{-0.3cm}

It should be noted that centralization defects indicate that certain sensitive operations are controlled by a single node, whose failure or compromise could lead to contract malfunction or attack. In contrast, existing defects usually involve situations where vulnerable functions can be directly accessed and exploited by others.

\noindent\textbf{(1) Mint Function with Single Signature (MFS):} 
This defect involves a function within the token contract that allows for the minting of an arbitrary number of tokens, and this function is controlled by a single-signature address, e.g., the contract creator. 

Typically, smart contract developers utilize the mint function to manage the stability of the tokens in the market. However, if the mint function is controlled by a single-signature address, it introduces a significant risk. Malicious project developers or attackers can exploit the mint function to generate an extensive quantity of tokens. This may lead to a significant influx of tokens in the market, ultimately resulting in substantial economic losses for the token holders.

\begin{figure}[h]
\setlength{\abovecaptionskip}{0.05cm}
\begin{lstlisting}[language=Solidity,mathescape]
contract Contract_Mint{
    mapping(address => uint256) private _balances;
    address public owner;
    modifier onlyOwner() {
        require(_msgSender() == _owner, "Only owner can perform this operation");
        _;
    }
    function mint(address account, uint256 amount) public onlyOwner{
        require(account != address(0), "ERC20: mint to the zero address");
        uint256 now_balances = _balances[account];
        _balances[account] = now_balances.add(amount);
    } }
\end{lstlisting}
\caption{An example of Mint Function with Single Signature defect.}
\label{fig:defect_mint}
\end{figure}

\textbf{Example:} 
A brief example of this defect is shown in Figure \ref{fig:defect_mint}.  
The \textit{mint} function (lines 8-12), controlled by a single-signature address via the \textit{onlyOwner} modifier (lines 4-7), enables the contract owner to mint tokens without restrictions.

\noindent\textbf{(2) Critical Variables Manipulation with Single Signature (CVS):} 
This defect is characterized by functions in smart contracts that can arbitrarily modify critical variables. If there are no restrictions on these functions and they can be invoked at will by a single-signature address, it may result in financial losses for users or even shut down the entire project. 

\textbf{Example:} 
In Figure \ref{fig:defect_crivar}, the function \textit{changeFee} allows the smart contract owner to arbitrarily adjust the transfer fee. As shown in the function \textit{transfer} (line 6-13), if an attacker, who has access as the owner, sets the transaction fee rate to 100\%, then all transferred amounts between users will be transferred to the \textit{\_owner} address as fees; consequently, the real recipient will not receive any token.

\vspace{-0.4cm}
\begin{figure}[h]
\setlength{\abovecaptionskip}{0.05cm}
\begin{lstlisting}[language=Solidity,mathescape]
contract Contract_CriVar{
    uint256 public _fee;
    function changeFee(uint256 newFee) public onlyOwner() {
        _fee = newFee;
    }
    function transfer(address to, uint256 amount) public virtual override returns (bool) {
        address from = _msgSender();
        uint256 actual_amount = amount * (100 - _fee) / 100;
        uint256 fee_amount = amount * _fee / 100;
        _transfer(from, to, actual_amount);
        _transfer(from, _owner, fee_amount);
        return true;
    } }
\end{lstlisting}
\caption{An example of Critical Variables Manipulation with Single Signature defect.}
\label{fig:defect_crivar}
\end{figure}
\vspace{-0.2cm}

\noindent\textbf{(3) Management without Timelock (MT):} 
This defect pertains to the ability of administrators to modify smart contracts without implementing a timelock mechanism. A timelock mechanism acts as a safeguard that ensures certain functions can only be executed after a predefined period. The absence of a timelock means that when smart contracts undergo critical changes or are under attack, users lack a time buffer to initiate necessary responses, such as withdrawal of their tokens. 

\vspace{-0.3cm}
\begin{figure}[h]
\setlength{\abovecaptionskip}{0.05cm}
\begin{lstlisting}[language=Solidity,mathescape]
modifier onlyAfter(uint256 time){
    if (block.timestamp <= time) revert TooEarly();
    _;}
modifier onlyBefore(uint256 time){
    if (block.timestamp >= time) revert TooLate();
    _;}
\end{lstlisting}
\caption{An example of timelock mechanism.}
\label{fig:defect_timelock}
\end{figure}
\vspace{-0.2cm}

\textbf{Example:} 
Figure \ref{fig:defect_timelock} shows an example of a Timelock mechanism implemented to prevent this defect. The timestamp of the current transaction is obtained from block.timestamp. Modifiers \textit{onlyAfter} (lines 1-4) and \textit{onlyBefore} (lines 5-8) check if the current time is after or before a predefined period. This mechanism ensures that functions using the modifier can only be executed before/after a certain time. 

\noindent\textbf{(4) Single Proxy Admin (SPA):} This defect concerns the utilization of a single-signature address to manage a proxy contract. 
A proxy contract is essentially a facilitator; users interact with it directly, and it is responsible for forwarding transactions to and from a secondary contract, known as the logic contract, which contains the actual operational logic~\cite{proxy}. Proxy contracts can be employed to upgrade smart contracts by modifying the address of the referenced logic contract.

However, if the authorization to modify the logical contract is held by a single-signature administrator address, this poses a security risk. These risks include losing the ability to upgrade the contract if the administrator's private key is lost, and the malicious developers or attackers can steal users' funds by altering the logic contract's address to one containing malicious functions.

\vspace{-0.4cm}
\begin{figure}[h]
\setlength{\abovecaptionskip}{0.05cm}
\begin{lstlisting}[language=Solidity,mathescape]
contract Contract_Proxy{
    modifier onlyAdmin() {
        require(_msgSender() == getAdmin(), "Only Admin can perform this operation");
        _;
    }
    function change_implementation(address new_implementation, bytes memory _data) public onlyAdmin{
        ERC1967Utils.upgradeToAndCall(new_implementation, _data);
    } }
\end{lstlisting}
\caption{An example of Single Proxy Admin defect.}
\label{fig:defect_proxy}
\end{figure}
\vspace{-0.3cm}

\textbf{Example:} 
The function \textit{change\_implementation} (lines 6-8) in Figure \ref{fig:defect_proxy} is employed to modify the address of the logic contract within the proxy contract. Due to the application of the \textit{onlyAdmin} modifier (lines 2-5), this function is exclusively invocable by a single-signature administrator address.

\noindent\textbf{(5) Self-destruct with Single Signature (SS):} 
This defect refers to the inclusion of the self-destruct operation that a single-signature address can trigger. In Ethereum, the self-destruct operation is employed to remove code from the blockchain and transfer the remaining Ethers (Ethereum's cryptocurrency) to a specified address~\cite{selfdestruct}. When the self-destruct operation can be triggered by a single-signature address, it poses the risk of the smart contract being terminated through a malicious invocation. 

\vspace{-0.4cm}
\begin{figure}[h]
\setlength{\abovecaptionskip}{0.05cm}
\begin{lstlisting}[language=Solidity,mathescape]
contract Contract_selfdestruct{
    function close() public {
        require(msg.sender == _owner, "Only the contract owner can call this function");
        selfdestruct(_owner);
    } }
\end{lstlisting}
\caption{An example of Self-destruct with Single Signature defect.}
\label{fig:defect_selfdestruct}
\end{figure}
\vspace{-0.3cm}

\textbf{Example:} 
An illustration of Self-destruct with Single Signature is shown in Figure \ref{fig:defect_selfdestruct}. A self-destruct operation (line 4) is within the function \textit{close}, which can be invoked by the \textit{\_owner} address. If the private key of the \textit{\_owner} address is compromised, an attacker can entirely destroy the smart contract at any time.

\noindent\textbf{(6) Individual Contract Output Reliance (IOR):} This defect arises when key variables or logic in a smart contract depend solely on the output of a single contract, such as a single smart contract oracle. Smart contract oracles are data feeds from external systems providing essential information to blockchains. If the single contract being relied upon fails or outputs incorrect information, it may lead to the execution of incorrect logic or incorrect parameters.

\vspace{-0.4cm}
\begin{figure}[h]
\setlength{\abovecaptionskip}{0.05cm}
\begin{lstlisting}[language=Solidity,mathescape]
contract Contract_Output{
    uint160 sqrtPrice = TickMath.getSqrtRatioAtTick(currentTick());
    uint256 price = FullMath.mulDiv(uint256(sqrtPrice).mul(uint256(sqrtPrice)), PRECISION, 2**(96*2));
    function currentTick() public view returns (int24 tick) {
        (, tick, , , , , ) = pool.slot0();
    } }
\end{lstlisting}
\caption{An example of Individual Contract Output Reliance defect.}
\label{fig:defect_output}
\end{figure}
\vspace{-0.3cm}

\textbf{Example:} As shown in Figure \ref{fig:defect_output}, the function \textit{currentTick} (line 4-6) directly retrieves the current price tick from a decentralized exchange \textit{pool} (line 5). Therefore, variable \textit{price} (line 3) directly relies on the output of contract \textit{pool}. There is no validation or any reference to the outputs from the contract. If the contract \textit{pool} fails or is attacked and outputs an incorrect result, then the token will be transferred at the wrong price.


\section{Methodology} \label{sec:method}

\subsection{Overview}

\begin{figure}[h]
    \centering
    \setlength{\abovecaptionskip}{0.05cm}
    \includegraphics[width=\linewidth]{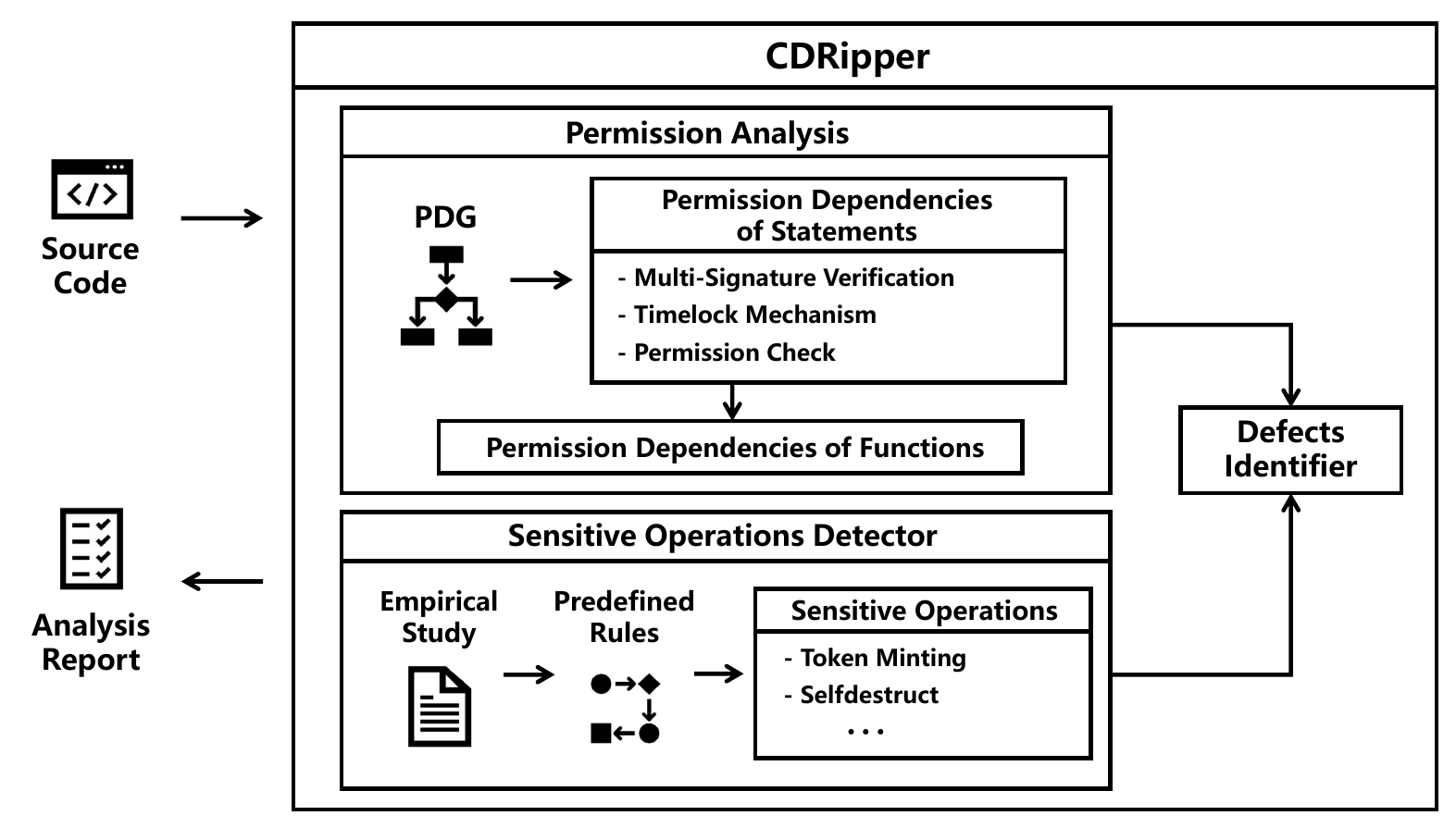}
    \caption{Overview of CDRipper}
    \label{fig:method_overview}
\end{figure}

The overview of CDRipper is depicted in Figure~\ref{fig:method_overview}. CDRipper consists of three primary modules: \textit{Permission Analysis}, \textit{Sensitive Operations Detector}, and \textit{Defects Identifier}. 

 \vspace{-0.4cm}
\vspace{-0.4cm}

CDRipper takes smart contracts' source code as its input. 
Firstly, the \textit{Permission Analysis} module constructs the Permission Dependency Graph (PDG) and extracts the permission dependencies of statements and functions, such as multi-signature verification, timelock mechanism, and permission check.
Subsequently, the \textit{Sensitive Operations Detector} identifies sensitive operations in functions based on the predefined rules summarized from the empirical study of centralization defects. 
Finally, CDRipper identifies centralization defects based on the permission dependencies and sensitive operations within the \textit{Defects Identifier} module.

\subsection{Permission Analysis}
Permission Analysis is a process that aims to identify the permission dependencies in smart contracts. To achieve this, we first construct a permission dependency graph (PDG), which captures statements that are constrained by permission. Next, the permission dependencies of functions can be extracted from the graph.

\subsubsection{Permission Dependency Graph (PDG)}
CDRipper constructs a Permission Dependency Graph (PDG) $\mathcal{G} = \left \{ N, E, \mathcal{X} \right \}$ to represent the permission dependencies of statements within smart contracts. This graph indicates whether the smart contract statements are constrained by permissions during execution, such as requiring certain privileges, multi-signature verification, or timelock. 

A node in a PDG represents a basic block of the Control Flow Graph (CFG). A basic block is a straight-line code sequence with no branches except at the entry and exit points~\cite{basicblock}. To analyze the permission dependency relationship between these basic blocks, CDRipper extracts the semantics of nodes and classifies them into four types.
An edge in a PDG represents the control flow dependency between two basic blocks. \textit{depend($b_1$, $b_2$)} is utilized to indicate that basic block $b_1$ is dependent on basic block $b_2$, which means that every execution path to basic block $b_1$ will pass through $b_2$.

$\mathcal{X}\left ( N \right ) \to \left \{ \mathsf{P},\mathsf{M},\mathsf{T},\mathsf{O}  \right \}$ is a labeling function that maps a node to one or more of the four types. A node \textit{n} is labeled as:
- $\mathsf{P}$ if the node \textit{n} is used to check whether the function caller is the high-privilege node,
- $\mathsf{M}$ if the node \textit{n} is used for multi-signature verification,
- $\mathsf{T}$ if the node \textit{n} is used as the timelock mechanism,
- $\mathsf{O}$ if the node \textit{n} is beyond the scope of the above three situations.

\vspace{-0.4cm}
\begin{table}[h] 
    \centering
    \setlength{\abovecaptionskip}{0.05cm}
        \caption{Fundamental Information of PDG.}
        \resizebox{\linewidth}{!}{
        \begin{tabular}{p{1.7cm}|p{6.3cm}}
            \hline
            \textbf{Name} & \textbf{Description} \\
            \hline
            \textit{depend($b_1$, $b_2$)} & Basic block $b_1$ is depend on basic block $b_2$ in CFG. \\
            \hline
            \textit{PerCheck(b)} & Basic block \textit{b} contains the logic to check whether the function caller is the high-privilege node. \\
            \hline
            \textit{MultiSig(b)} & Basic block \textit{b} contains the multi-signature verification. \\
            \hline
            \textit{Timelock(b)} & Basic block \textit{b} contains the timelock mechanism. \\
            \hline
        \end{tabular}}
    \label{tab:method_per_pdg}
\end{table}
 \vspace{-0.3cm}

CDRipper uses Slither~\cite{feist2019slither}, a static analysis framework, to help construct the PDG. The fundamental information of the PDG is detailed in Table~\ref{tab:method_per_pdg}.

\textit{MultiSig(b)} is utilized to verify multi-signatures within the basic block \textit{b}. This verification process requires two key elements: the signature information provided by multiple private keys and a threshold indicating the minimum number of signatures for successful verification. Typically, multi-signature information is obtained from the input of the contract caller, while the threshold can be obtained from user input or a predefined state variable. As a result, we establish the following criteria to identify the presence of multi-signature verification within a basic block: (i) the existence of multi-signature information in the user's input, (ii) reading an integer type value as the threshold, and (iii) utilizing both the multi-signature information and the threshold value in the same conditional logic within the basic block.

\textit{Timelock(b)} is utilized to identify a timelock mechanism within the basic block \textit{b}. Implementing a timelock mechanism in smart contracts requires the current timestamp of the transaction and a pre-defined time threshold to determine unlocking eligibility. Therefore, we consider a basic block to have a timelock mechanism if: (i) the basic block contains statements to retrieve the current transaction timestamp, such as using ``\textit{block.timestamp}''. (ii) the basic block accesses a time threshold. (iii) both the current transaction timestamp and the time designated for timelock unlocking are used within the same conditional logic within the basic block.

\textit{PerCheck(b)} is utilized to determine whether basic block \textit{b} contains logic to check whether the function caller is the high-privilege node.
The implementation of permission check logic in smart contracts requires the address of the function caller and a pre-defined set of high-privilege nodes. Our analysis indicates that high-privilege nodes are typically set during contract creation. These nodes may not necessarily be stored as variables of the address type; for example, a $mapping \left ( address \Rightarrow bool \right ) $ might be used to determine whether an address is a high-privilege node. Therefore, we consider a basic block to have permission check logic if: (i) the block includes statements to retrieve the address of the function caller, such as using \textit{msg.sender}, (ii) the block accesses the set of high-privilege nodes, and (iii) both the current function caller and the set of high-privilege nodes are used within the same conditional logic within the block.

\subsubsection{Permission Dependencies of Functions}
CDRipper identifies the permission dependencies of functions by analyzing the PDG. This analysis is based on two main criteria. Firstly, the function itself has permission restrictions. Secondly, the function calls other functions with permission restrictions. The rules for extracting the permission dependencies of functions are illustrated in Figure~\ref{fig:method_per_func}.

\vspace{-0.4cm}
\begin{figure}[h] 
    \centering
    \setlength{\abovecaptionskip}{0.05cm}
    
    \begin{equation}
    \resizebox{0.7\hsize}{!}{
    $MultiSig\left ( f \right ) \gets \exists b \in Block(f),\ MultiSig(b)$}
    \label{eq:method_per_func_multisig_1}
    \end{equation}
    \begin{equation}
    \resizebox{0.6\hsize}{!}{
    $MultiSig(f) \gets Call(f, \bar{f} ),\ MultiSig(\bar{f})$}
    \label{eq:method_per_func_multisig_2}
    \end{equation}
    \begin{equation}
    \resizebox{0.7\hsize}{!}{
    $Timelock\left ( f \right ) \gets \exists b \in Block(f),\ Timelock(b)$}
    \label{eq:method_per_func_time_1}
    \end{equation}
    \begin{equation}
    \resizebox{0.6\hsize}{!}{
    $Timelock(f) \gets Call(f, \bar{f} ),\ Timelock(\bar{f})$}
    \label{eq:method_per_func_time_2}
    \end{equation}
    \begin{equation}
    \resizebox{0.9\hsize}{!}{
    $Limited(f) \gets \exists b_1 \in Block(f),\ depend(b_1, b_2),\ PerCheck(b_2) $}
    \label{eq:method_per_func_per_1}
    \end{equation}
    \begin{equation}
    \resizebox{0.6\hsize}{!}{
    $Limited(f) \gets Call(f, \bar{f}),\ Limited(\bar{f})$}
    \label{eq:method_per_func_per_2}
    \end{equation}
    \begin{equation}
    \resizebox{0.7\hsize}{!}{
    $LimitedPublic(f) \gets Limited(f),\ Public(f)$}
    \label{eq:method_per_func_per_3}
    \end{equation}
    \caption{Rules for extracting the permission dependencies of functions}
    \label{fig:method_per_func}
\end{figure}
\vspace{-0.2cm}

\textit{MultiSig(f)} and \textit{Timelock(f)} indicate that function \textit{f} contains logic that can only be executed after being verified by a multi-signature or unlocked by a timelock, respectively. 
Meanwhile, \textit{Limited(f)} and \textit{LimitedPublic(f)} signify that the function \textit{f} can only be invoked by a high-privilege node. In addition, \textit{LimitedPublic(f)} further requires that the function \textit{f} needs to be callable directly, potentially serving as an entry point for an attacker to launch an attack.

\subsection{Sensitive Operations Detector}
Sensitive Operations Detector is designed to identify sensitive operations related to centralization defects in smart contracts. These operations are identified based on the empirical study outlined in Section~\ref{sec:defects}. Specifically, the sensitive operations related to centralization defects are detailed in Table~\ref{tab:method_opera}.

\vspace{-0.4cm}
\begin{table}[h] 
    \centering
    \setlength{\abovecaptionskip}{0.05cm}
        \caption{Sensitive Operations related to centralization defects.}
        \resizebox{\linewidth}{!}{
        \begin{tabular}{p{2.5cm}|p{6cm}}
            \hline
            \textbf{Name} & \textbf{Description} \\
            \hline
            \textit{Mint(func)} & Function \textit{func} is designated for token minting. \\
            \hline
            \textit{ModifyCriVar(func)} & Function \textit{func} is employed to modify the value of critical variables. \\
            \hline
            \textit{ChangeImple(func)} & Function \textit{func} is utilized to change the implementation address of proxy contracts. \\
            \hline
            \textit{self-destruct(func)} & Function \textit{func} contains the self-destruct operation. \\
            \hline
            \textit{Token(C)} & Contract \textit{C} is a token contract. \\
            \hline
            \textit{Proxy(C)} & Contract \textit{C} is a proxy contract. \\
            \hline
            \textit{dependOutput(var, out)} & The value of variable \textit{var} is depend on variable \textit{out}, which is the output of other contracts. \\
            \hline
        \end{tabular}}
    \label{tab:method_opera}
\end{table}
\vspace{-0.3cm}

\textit{Mint(func)} indicates that the function \textit{func} is used for token minting. A function is considered a token minting function if it includes the logic to increase or modify token balances without reducing the balances of any address. To identify \textit{Mint Function with Single Signature} defect, we specifically examine token contracts using \textit{Token(C)} to check if contract \textit{C} is a token contract. This approach allows us to exclude non-token contracts and concentrate on the potential functions for token minting, thus improving efficiency and accuracy.

\textit{ModifyCriVar(func)} signifies that function \textit{func} is utilized to modify the value of critical variables in smart contracts.
We observed that critical variables are usually defined and stored in the EVM storage during the contract’s creation. 
Therefore, CDRipper identifies \textit{ModifyCriVar(func)} by subsequently identifying these critical variables and the corresponding functions responsible for writing to their storage.

\textit{Proxy(C)} signifies that contract \textit{C} is a proxy contract, and \textit{ChangeImple(func)} signifies that function \textit{func} is utilized to change the implementation address of proxy contracts. We utilized the USCHunt~\cite{bodell2023proxy} tool to determine if the target contract is a proxy and to identify the address variables of the logic contract.

\textit{self-destruct(func)} signifies that function \textit{func} contains the self-destruct operation. Since self-destruct is a special operation in EVM, we employ the keyword to identify it. 

\textit{dependOutput(var, out)} signifies that state variables \textit{var} rely on the output of external calls to other contracts. The variable \textit{out} is defined by the output of external calls in other contracts, and it is located by analyzing the \textit{CALL} and \textit{DELEGATECALL} operations. Then, we identify the state variable \textit{var} whose value depends on the value of \textit{out} by analyzing the data flow graph of smart contracts.

\subsection{Defects Identifier}
The \textit{Defects Identifier} module utilizes the permission dependencies of functions and the sensitive operations extracted from the previous modules to identify centralization defects. We will provide more details on how we detect centralization defects in the following paragraphs.


\noindent\textbf{(1) Mint Function with Single Signature (MFS)}
Detecting \textit{Mint Function with Single Signature} involves two primary criteria. Firstly, the existence of a token minting function within the smart contract; secondly, this function is controlled by a single-signature address. The rule for detecting Mint Function with Single Signature is shown in Formula~\ref{equa:method_identifer_mint}.

\begin{equation}
\resizebox{0.7\hsize}{!}{
$\displaystyle\frac{\begin{array}{c}
            Token(C),\ \exists f\in Func(C),\ Mint(f),\\ LimitedPublic(f),\ 
            \neg MultiSig(f)
          \end{array}}{Mint\ Function\ with\ Single\ Signature}$}
\label{equa:method_identifer_mint}
\end{equation}

\noindent\textbf{(2) Critical Variables Manipulation with Single Signature (CVS)}
Detecting this defect involves two primary criteria. Firstly, the existence of a function to modify the critical variables in smart contracts. Secondly, this function is controlled by a single-signature address. 
The rule for detecting Critical Variables Manipulation with Single Signature is shown in Formula \ref{equa:method_identifer_crivar}.
\begin{equation}
\resizebox{\hsize}{!}{
$\displaystyle\frac{\begin{array}{c}
            \exists f\in Func(C),\ ModifyCriVar(f), \\
            LimitedPublic(f),\ \neg MultiSig(f) 
          \end{array}}{Critical\ Variables\ Manipulation\ with\ Single\ Signature}$}
\label{equa:method_identifer_crivar}
\end{equation}

\noindent\textbf{(3) Management without Timelock (MT)} 
Detecting the \textit{Management without Timelock} defect involves two primary criteria. Firstly, the existence of a management function in the smart contract. Secondly, the absence of a timelock mechanism within this function. The rule for detecting Management without Timelock is shown in Formula \ref{equa:method_identifer_timelock}.
\begin{equation}
\resizebox{0.8\hsize}{!}{
$\displaystyle\frac{\begin{array}{c}
             \exists f\in Func(C),\ ModifyCriVar(f, var), \\
            LimitedPublic(f),\ \neg Timelock(f) 
          \end{array}}{Management\ without\ Timelock}$}
\label{equa:method_identifer_timelock}
\end{equation}

\noindent\textbf{(4) Single Proxy Admin (SPA)}
Detecting Single Proxy Admin involves three primary criteria. 
Firstly, the contract should be a proxy contract. Secondly, it should have a high-privilege address with functions to modify the logic contract's address. Thirdly, the high-privilege address should be a single-signature address. 
The rule for detecting Single Proxy Admin is shown in Formula \ref{equa:method_identifer_proxy}.
\begin{equation}
\resizebox{\hsize}{!}{
$\displaystyle\frac{\begin{array}{c}
            Proxy(C),\ \exists f\in Func(C),\ ChangeImple(f),\ \neg MultiSig(f) 
          \end{array}}{Single\ Proxy\ Admin}$}
\label{equa:method_identifer_proxy}
\end{equation}

\noindent\textbf{(5) Self-destruct with Single Signature (SS)}
Two primary criteria are used to detect the SS defect, i.e.,  the presence of the self-destruct operation and its exclusive invocation by a single-signature address. 
The rule for detecting Self-destruct with Single Signature is shown in Formula \ref{equa:method_identifer_self-destruct}.
\begin{equation}
\resizebox{0.65\hsize}{!}{
$\displaystyle\frac{\begin{array}{c}
            \exists f\in Func(C),\ self-destruct(f),\\
            LimitedPublic(f), \
            \neg MultiSig(f)
          \end{array}}{Self-destruct\ with\ Single\ Signature}$}
\label{equa:method_identifer_self-destruct}
\end{equation}

\noindent\textbf{(6) Individual Contract Output Reliance (IOR): }
The primary criterion for this defect is whether the state variables in the contract rely solely on the output of individual external calls to other contracts.
The rule for detecting Individual Contract Output Reliance is shown in Formula \ref{equa:method_identifer_output}.
\begin{equation}
\resizebox{0.9\hsize}{!}{
$\displaystyle\frac{\begin{array}{c}
             \exists var \in StateVar(C),\ dependOutput(var, out_1), \\ 
            \nexists dependOutput(var, out_2),\ out_1 \ne out_2\  
          \end{array}}{Individual\ Contract\ Output\ Reliance}$}
\label{equa:method_identifer_output}
\end{equation}
\vspace{-0.4cm}

\section{Evaluation} \label{sec:exp}

\subsection{Experimental Setup}
The experiment was executed on a server running Ubuntu 22.04.1 LTS equipped with 70 Intel(R) Xeon(R) Platinum 8360H CPUs (3.00GHz) and 80 GB of memory.

\noindent\textbf{Dataset.}
To identify centralization defects in real-world Ethereum smart contracts, we utilized an open-source dataset from a GitHub repository~\cite{dataset}. This dataset contains the source code of all verified smart contracts on Etherscan, which we downloaded in November 2023. Given that Solidity 0.8.21 was the latest compiler version at the time of writing, we chose this version and excluded contracts that failed to compile. This process yielded 323,329 verified smart contracts. 

Subsequently, we conducted preprocessing on the dataset, involving two primary steps: contract deduplication and the exclusion of toy contracts. 
Initially, duplicates were identified by comparing the source code of the smart contracts~\cite{sendner2024large}, which led to the removal of 15,133 duplicate smart contracts. Contracts with fewer than two transactions were then excluded to eliminate toy contracts, resulting in the exclusion of an additional 63,772 contracts. This refinement process ultimately yielded a dataset of 244,424 smart contracts.

\vspace{-0.2cm}
\begin{table}[h]
    \centering
    \setlength{\abovecaptionskip}{0.05cm}
    \caption{The Statistics of Dataset}
    \resizebox{\linewidth}{!}{
        \begin{tabular}{|l|r|r|}
        \hline
        \multicolumn{1}{|c|}{\textbf{Dataset}} & \multicolumn{1}{c|}{\textbf{\# Trans (Avg)}} & \multicolumn{1}{c|}{\textbf{\# LoC (Avg)}} \\ \hline
        Preprocessed Dataset               & 421.39                                       & 630.38                                     \\ \hline
        Smart Contracts with Defects           & 630.38                                       & 783.23                                     \\ \hline
        \end{tabular}
    }
    \label{tab:data_stat}
\end{table}
\vspace{-0.2cm}

The statistics of the dataset are presented in Table~\ref{tab:data_stat}. In the preprocessed dataset, the average number of transaction records per contract is 421.39, and the average number of lines of code is 630.38.


\noindent\textbf{Evaluation Metrics.}
We summarize the following research questions (RQ) to evaluate the effectiveness of CDRipper.
\begin{itemize}
\item RQ1: What are the prevalent of the six centralization defects in Ethereum?
\item RQ2: In terms of effectiveness, how is the precision of CDRipper in finding centralization defects in smart contracts?
\end{itemize}

\subsection{Answer to RQ1: Prevalence of Defects}
To answer RQ1, we executed CDRipper on 244,424 verified Ethereum smart contracts. The corresponding experimental results are presented in Table \ref{tab:exp_detection}. The second and third columns of the table display the number and proportion of various defects detected by CDRipper, respectively. As CDRipper primarily aims to determine whether there exist centralization defects within contracts, instances of the same defect occurring multiple times in a single contract are considered only once.

CDRipper identified 82,446 smart contracts containing at least one centralization defect, representing 33.7\%. \textit{Critical Variables Manipulation with Single Signature (CVS)} and \textit{Management without Timelock (MT)} are the two most prevalent centralization defects, present in approximately 30.28\% and 30.34\% of smart contracts, respectively.
In contrast, \textit{Self-destruct with Single Signature (SS)} exhibited the lowest frequency, occurring at a rate of merely 0.62\%. The frequencies of other centralization defect types are relatively consistent, falling within the range of 1\% to 3\%.
\vspace{-0.4cm}
\begin{table}[h]
    \centering
    \setlength{\abovecaptionskip}{0.05cm}
    \caption{The defects detected by CDRipper}
    \resizebox{\linewidth}{!}{
        \begin{tabular}{|l|r|r|}
        \hline
        \multicolumn{1}{|c|}{\textbf{Centralization Defects}}                                                  & \multicolumn{1}{c|}{\textbf{\# Defects}} & \multicolumn{1}{c|}{\textbf{Per(\%)}} \\ \hline
        Mint Function Control by Single Signature (MFS)                                                        & 5,947                                    & 2.43                                   \\ \hline
        \begin{tabular}[c]{@{}l@{}}Critical Variables Manipulation with Single \\ Signature (CVS)\end{tabular} & 74,008                                   & 30.28                                  \\ \hline
        Management without Timelock (MT)                                                                       & 74,154                                   & 30.34                                  \\ \hline
        Single Proxy Admin (SPA)                                                                               & 3,422                                    & 1.40                                   \\ \hline
        Self-destruct with Single Signature (SS)                                                                & 1,525                                    & 0.62                                   \\ \hline
        Individual Contract Output Reliance (IOR)                                                              & 2,443                                    & 1.00                                   \\ \hline
        \multicolumn{1}{|c|}{\textbf{All}}                                                                     & 82,446                                   & 33.73                                  \\ \hline
        \end{tabular}
    }
    \label{tab:exp_detection}
\end{table}
\vspace{-0.3cm}

\noindent{\bf Analysis of the Prevalence.} 
Our analysis revealed that 33.7\% of the contracts examined exhibited at least one form of centralization defect, which is notably high. These defects are predominantly attributed to \textit{CVS} and \textit{MT}. The high prevalence of defects can largely be attributed to DeFi developers overlooking them. Unlike traditional vulnerabilities that can result in direct financial losses, these defects may sometimes be harmless to developers and even offer certain advantages, such as allowing contract owners to alter critical variables like fees. Additionally, addressing these defects could increase development costs, leading some developers to retain them. However, these defects can harm users and erode their trust in the contracts, as discussed in Section~\ref{sec:definition}.

As shown in Table~\ref{tab:data_stat}, among the contracts identified by CDRipper as having centralization defects, the average number of transactions and lines of code are 630.38 and 783.23, respectively. Notably, the average number of transactions represents a significant increase of 49.6\% compared to the original dataset. This indicates that filtering the dataset based on the number of transactions can enhance its quality.

\subsection{Answer to RQ2: Evaluation of CDRipper}
To answer RQ2, we evaluate the performance of CDRipper in this subsection. 
Due to the infrequent occurrence of certain centralization defects, directly sampling the entire dataset may not capture a sufficient number of these defects, making it challenging to accurately evaluate detection effectiveness. For instance, as demonstrated in Table~\ref{tab:exp_detection}, contracts containing the \textit{Self-destruct with Single Signature} defect comprise only 0.62\% of the total. Consequently, direct sampling may yield a sample devoid of any contracts with this defect, thus impeding the verification of CDRipper’s accuracy in its detection.
To address this, we conducted random sampling validation respectively for CDRipper’s precision in detecting various types of centralization defects and for its false negative rate.

To validate the precision of CDRipper, we conduct a random sampling on smart contracts identified as positive by CDRipper. We employ the sampling approach based on a 95\% confidence level~\cite{confidence} and a 10\% confidence interval~\cite{samplesizecal}, aligned with previous studies~\cite{yang2023definition, jiang2018contractfuzzer, kalra2018zeus, luu2016making}. Two researchers manually verified the detection results, recording true positives (TP) and false positives (FP) to analyze CDRipper's performance. 

The results are presented in Table \ref{tab:exp_prec}. The second column illustrates the sample quantities for each type of centralization defect. In the third and fourth columns, we provide the counts of true positives (TP) and false positives (FP), respectively. The fifth column displays the precision of CDRipper in identifying various defects, calculated using the formula $\frac{\#TP}{\#TP + \#FP} \times 100\%$. Additionally, we computed the overall precision to gauge the effectiveness of CDRipper. The overall precision is determined by $ \frac{{\textstyle \sum_{i=1}^{n}p_{c_i} \times \left | c_i \right | }}{ {\textstyle \sum_{i=1}^{n} \left | c_i \right | } } $, where $p_{c_i}$ represents the precision of detecting defect $i$, and $\left | c_i \right |$ is the number of smart contracts identified with centralization defect $i$.

\vspace{-0.3cm}
\begin{table}[h]
    \centering
    \setlength{\abovecaptionskip}{0.05cm}
    \caption{The precision of CDRipper}
    \resizebox{\linewidth}{!}{
        \begin{tabular}{|l|r|r|r|r|}
        \hline
        \multicolumn{1}{|c|}{\textbf{Centralization Defects}}                                                 & \multicolumn{1}{c|}{\textbf{\# Sam}} & \multicolumn{1}{l|}{\textbf{\# TP}} & \multicolumn{1}{l|}{\textbf{\# FP}} & \multicolumn{1}{c|}{\textbf{Prec(\%)}} \\ \hline
        Mint Function Control by Single Signature (MFS)                                                       & 94                                       & 86                                  & 8                                   & 91.5                                         \\ \hline
        \begin{tabular}[c]{@{}l@{}}Critical Variables Manipulation with Single\\ Signature (CVS)\end{tabular} & 95                                       & 92                                  & 3                                   & 96.8                                         \\ \hline
        Management without Timelock (MT)                                                                      & 95                                       & 93                                  & 2                                   & 97.9                                         \\ \hline
        Single Proxy Admin (SPA)                                                                              & 93                                       & 85                                  & 8                                   & 91.4                                         \\ \hline
        Self-destruct with Single Signature (SS)                                                               & 90                                       & 90                                  & 0                                   & 100                                          \\ \hline
        Individual Contract Output Reliance (IOR)                                                             & 92                                       & 78                                  & 14                                  & 84.8                                         \\ \hline
        \multicolumn{1}{|c|}{\textbf{All}}                                                                    & \textbackslash{}                         & \textbackslash{}                    & \textbackslash{}                    & 93.7                                         \\ \hline
        \end{tabular}
    }
    \label{tab:exp_prec}
\end{table}
\vspace{-0.3cm}

The identification of smart contracts incorporating the \textit{Self-destruct with Single Signature} obtains a precision rate of 100\%. For \textit{Mint Function with Single Signature}, \textit{Critical Variables Manipulation with Single Signature}, \textit{Management without Timelock}, \textit{Single Proxy Admin}, \textit{Individual Contract Output Reliance}, CDRipper reports them at a precision of 91.5\%, 96.8\%, 97.9\%, 91.4\%, 84.8\%, respectively. Moreover, CDRipper demonstrates an overall precision of 93.7\%.

\noindent\textbf{False Positives.}
The experimental findings reveal instances of false positives. In the context of the \textit{Mint Function Controlled by Single Signature}, among the 94 sampled contracts, 8 are false positives due to the misidentification of variables of token balances. 

\vspace{-0.4cm}
\begin{figure}[h]
\setlength{\abovecaptionskip}{0.06cm}
\begin{lstlisting}[language=Solidity,mathescape]
mapping(address => uint256) public registeredContracts; // 0 EMPTY, 1 ERC1155, 2 ERC721, 3 HANDLER, 4 ERC20, 5 BALANCE, 6 CLAIM, 7 UNKNOWN, 8 FACTORY, 9 STAKING, 10 BYPASS
function registerContract(address _contract, uint _type) public isRegisteredContractOrOwner(_msgSender()) {
        registeredContracts[_contract] = _type;
        registeredOfType[_type].push(_contract);
    }
\end{lstlisting}
\caption{An example of false positive detected by CDRipper.}
\label{fig:exp_fp}
\end{figure}
\vspace{-0.3cm}

Illustrated in Figure \ref{fig:exp_fp} is an example of a false positive\cite{fp_con}, where the variable \textit{registeredContract} (line 1) functions as a mapping from contract addresses to their respective types. Remarkably, the data structure of this variable is the same as that of the token balances in ERC20 tokens~\cite{erc20}. Consequently, CDRipper interprets the function \textit{registerContract} (line 2-5) as one capable of arbitrarily modifying token balances, thereby characterizing it as a mint function.

In the case of \textit{Critial Variables Manipulation with Single Signature}, certain functions initialize critical variables before reading them, leading CDRipper to identify these functions as modifying crucial variables, resulting in false positives. For \textit{Management without Timelock}, false positives arise due to the incorrect identification of variables intended for contract management. Regarding \textit{Single Proxy Admin}, false positives are due to misjudgments concerning proxy contract behavior. 
As for \textit{Individual Contract Output Reliance}, false positives stem from misinterpretations of external function calls.

\noindent\textbf{False Negatives.}
To find contracts with centralization defects that CDRipper failed to report, we follow the same sampling method used for precision analysis. We randomly sampled 96 contracts from 161,978 contracts where no defect was reported, using a confidence interval of 10 and a confidence level of 95\%. We then manually label them to find false negatives that CDRipper missed. The number of false negatives of CDRipper is shown in Table \ref{tab:exp_fn}. In total, we find that 14 of 96 contracts are false negatives; the false negative rate of CDRipper is 14.6\%.

\vspace{-0.4cm}
\begin{table}[h]
    \centering
    \setlength{\abovecaptionskip}{0.05cm}
    \caption{The number of false negatives of CDRipper}
    \resizebox{\linewidth}{!}{
        \begin{tabular}{|l|r|r|}
        \hline
        \multicolumn{1}{|c|}{\textbf{Centralization Defects}}                                                 & \multicolumn{1}{c|}{\textbf{\# FN}} & \multicolumn{1}{c|}{\textbf{Per (\%)}} \\ \hline
        Mint Function Control by Single Signature (MFS)                                                       & 9                                   & 9.4                                    \\ \hline
        \begin{tabular}[c]{@{}l@{}}Critical Variables Manipulation with Single\\ Signature (CVS)\end{tabular} & 7                                   & 7.3                                    \\ \hline
        Management without Timelock (MT)                                                                      & 7                                   & 7.3                                    \\ \hline
        Single Proxy Admin (SPA)                                                                              & 1                                   & 1                                      \\ \hline
        \multicolumn{1}{|c|}{\textbf{All}}                                                                    & 14                                  & 14.6                                   \\ \hline
        \end{tabular}}
    \label{tab:exp_fn}
\end{table}
\vspace{-0.3cm}

False negatives in the \textit{Mint Function with Single Signature} result from the failure to discern certain complex data structures associated with non-fungible tokens (NFT), leading to the inability to identify the corresponding NFT minting function. 
The false negatives of \textit{Critical Variables Manipulation with Single Signature} and \textit{Management without Timelock} result from the failure to discern certain complex permission mechanisms.
Specifically, some contracts utilize the data structure $mapping \left ( bytes32 \Rightarrow RoleData \right ) $ to store permission information. In this structure, \textit{bytes32} corresponds to string types representing high-permission nodes (e.g., \textit{DEFAULT\_ADMIN\_ROLE}), and \textit{RoleData} is a custom data structure. The complexity of this custom data structure poses challenges for CDRipper in accurately categorizing it as a permission-related variable.
The false negatives related to \textit{Single Proxy Admin} arise because CDRRipper fails to identify this contract as a proxy contract. 

\section{Discussion} \label{sec:disc}

\subsection{Cross-Contract Analysis}
The implementation of multi-signature verification and timelock mechanisms can be accomplished by transferring contract permissions to a multi-signature wallet contract or a timelock contract. This process involves the transfer of permissions between distinct contracts, necessitating transaction records for analysis. As CDRipper solely utilizes smart contract source code as its input, it lacks the capability to recognize this scenario, resulting in false positives. To quantify the number of reported false positives in this context, we have devised an algorithm to conduct cross-contract analysis. The central aspect of the algorithm involves identifying the transfer of ownership through the examination of transaction records.

Table \ref{tab:disc_fp} shows the number of false positives that occur when cross-contract analysis is omitted. All false positive rates are less than 2\%. This indicates that while the concern of centralization defects is important, developers seldom address this issue by transferring contract permissions to external multi-signature wallet contracts and timelock contracts.

\vspace{-0.3cm}
\begin{table}[h]
    \centering
    \setlength{\abovecaptionskip}{0.05cm}
    \caption{The number of false positives resulting from the omission of cross-contract analysis}
    \resizebox{\linewidth}{!}{
        \begin{tabular}{|l|r|r|}
        \hline
        \multicolumn{1}{|c|}{\textbf{Centralization Defects}}                                                 & \multicolumn{1}{c|}{\textbf{\# FP}} & \multicolumn{1}{c|}{\textbf{Per(\%)}} \\ \hline
        Mint Function Control by Single Signature (MFS)                                                       & 43                                  & 0.72                                  \\ \hline
        \begin{tabular}[c]{@{}l@{}}Critical Variables Manipulation with Single\\ Signature (CVS)\end{tabular} & 200                                 & 0.27                                  \\ \hline
        Management without Timelock (MT)                                                                      & 684                                 & 0.92                                  \\ \hline
        Single Proxy Admin (SPA)                                                                              & 63                                  & 1.84                                  \\ \hline
        Self-destruct with Single Signature (SS)                                                               & 3                                   & 0.20                                  \\ \hline
        \end{tabular}}
    \label{tab:disc_fp}
\end{table}
\vspace{-0.2cm}

\subsection{Possible Solution for Centralization Defects}
To mitigate the risk of centralization defects in smart contracts, we have developed CDRipper specifically for detecting such issues. Furthermore, our objective is to support smart contract developers in creating secure smart contracts. In this subsection, we provide possible solutions to assist developers in avoiding the identified centralization defects.

 \vspace{-0.4cm}
\begin{table}[h] 
    \centering
    \setlength{\abovecaptionskip}{0.05cm}
        \caption{Possible solutions for the centralization defects.}
        \resizebox{\linewidth}{!}{
        \begin{tabular}{p{3.6cm}|p{5.3cm}}
            \hline
            \textbf{Centralization Defect} & \textbf{Possible Solutions} \\
            \hline
            \textit{Mint Function with Single Signature (MFS)} & Eliminate the mint function or implement multi-signature verification prior to minting. \\
            \hline
            \textit{Management without Timelock (MT)} & Implement timelock mechanism prior to the execution of management functions. \\
            \hline
            \textit{Critical Variables Manipulation with Single Signature (CVS)} & Implement multi-signature verification prior to modifying critical variables.\\
            \hline
            \textit{Single Proxy Admin (SPA)} & Transfer administrative permissions of a proxy contract to a multi-signature wallet contract. \\
            \hline
            \textit{Self-destruct with Single Signature (SS)} & Eliminate self-destruct or implement multi-signature verification prior to self-destruct. \\
            \hline
            \textit{Individual Contract Output Reliance (IOR)} & Implement verification of contract output or combine multiple contract outputs. \\
            \hline
        \end{tabular}}
    \label{tab:disc_sol}
\end{table}
\vspace{-0.2cm}

Table \ref{tab:disc_sol} presents concise potential solutions for each defect. 
To address centralization defects related to sensitive functions controlled by a single signature, such as \textit{Mint Function with Single Signature}, \textit{Critical Variables Manipulation with Single Signature}, \textit{Single Proxy Admin}, and \textit{Self-destruct with Single Signature}, developers should consider removing these sensitive functions from smart contracts. If these functions are essential for the project, we recommend restricting them by multi-signature verification. There are two approaches to multi-signature verification: direct verification before function execution, and transferring control of the contract to a multi-signature wallet contract. It is important to note that while multi-signature authentication can help mitigate centralization defects, it cannot completely eliminate them. There is still a risk of multiple private keys being stolen~\cite{lamby2023trusting}.

For \textit{Management Without Timelock}, we suggest developers restrict management functions using a timelock mechanism. There are two approaches to a timelock mechanism: direct timelock verification before function execution and transfer control of the contract to a timelock contract.

Regarding \textit{Individual Contract Output Reliance}, developers have two viable solutions: one involves the verification of whether the contract output aligns with expected criteria, while the other entails considering outputs from various distinct contracts or outputs from the same contract at different time.


\subsection{Limitations}
First, CDRipper uses smart contract source code as its input, which makes it unable to identify centralization defects in unverified smart contracts. However, compared to bytecode, source code allows for a more accurate analysis of permissions dependencies in the contract, resulting in overall improved accuracy. Additionally, the source-based approach is already sufficient to meet developers' needs in detecting centralization defects. 
Therefore, we have decided to use source code as the input for CDRipper.


Besides, CDRipper’s current functionality is limited to detecting six specific defect types, which are identified based on existing posts and audit reports. 
It is possible that the current source may overlook some defects.
However, our objective is to ensure that all defined defects have a real-world source to guarantee their reliability. In the future, as new posts or reports emerge, we plan to use the same open card sorting method to create new cards and categorize new defects accordingly.

Thirdly, CDRipper’s detection rules rely on expert knowledge of smart contracts and their centralization defects, which complicates updates and raises barriers to use. Future work could involve developing an automated method for generating detection rules to lower usage barriers and simplify updates, enabling CDRipper to detect newly emerging centralization vulnerabilities more swiftly.



\section{Related Work} \label{sec:related_work}
\subsection{Detection of Security Problems in Smart Contracts} 
In recent years, researchers have extensively explored the topic of defects in smart contracts. Chen et al.~\cite{chen2020defining} were pioneers in defining and categorizing smart contract defects. By analyzing posts about smart contracts on Ethereum Stack Exchange, they identified 20 distinct types of smart contract defects. Subsequently, they developed DefectChecker~\cite{chen2021defectchecker}, a tool aimed at detecting these identified defects. Yang et al.~\cite{yang2023definition} focused on defects in NFT smart contracts, defining five distinct types of NFT defects. They introduced NFTGuard, a tool specifically designed to detect these NFT defects.

Furthermore, several researchers have conducted studies on vulnerabilities in smart contracts. Much research employs static analysis methods for the analysis and detection of vulnerabilities in smart contracts. For example, ONENTE~\cite{luu2016making}, ZEUS~\cite{kalra2018zeus}, GASPER~\cite{chen2017under}, Slithercite~\cite{feist2019slither}, Securify~\cite{tsankov2018securify}, Ethainter~\cite{brent2020ethainter}, and Manian~\cite{nikolic2018finding}. OYENTE stands out as the pioneering work in smart contract vulnerability detection, conducting a comprehensive exploration of security vulnerabilities and developing a detection tool based on symbolic execution~\cite{luu2016making}. Other works, such as Contract-Fuzzer~\cite{jiang2018contractfuzzer}, ETHBMCcite~\cite{frank2020ethbmc}, Mythrilcite{durieux2020empirical}, and Echidna~\cite{grieco2020echidna}, utilize dynamic analysis approaches. Additionally, Ethircite~\cite{albert2018ethir}, KEVM~\cite{hildenbrandt2018kevm}, and Isabelle~\cite{amani2018towards} leverage formal verification methods.

However, these tools face challenges in detecting centralization defects in smart contracts due to a lack of analysis on centralized permission issues.

\subsection{DeFi Centralization Security Risks Analysis}
In response to the escalating economic losses attributed to centralization security risks, an increasing number of researchers are making efforts to address these concerns. This research can be categorized into three main types: the detection of smart contract backdoors, the identification of Rug Pull scams, and the detection of privileged nodes.

Smart contract backdoor detection has been addressed in works such as BadApple~\cite{yan2023bad} and Pied-Piper~\cite{ma2023pied}. Yan et al.~\cite{yan2023bad} focus on centralization risks in cryptocurrency wallets and DApps, presenting seven patterns and a specialized detection algorithm for Android cryptocurrency wallets. Ma et al.~\cite{ma2023pied} identify five backdoors in smart contracts, proposing a detection method that integrates datalog and fuzzing. Lamby et al.~\cite{lamby2023trusting} define centralization risk as the source code containing privileged access patterns on fund-modifying logic.
However, these works have a narrow focus, ignoring other potential forms of centralization defects beyond backdoors, and defects in other types of contracts besides token contracts or cryptocurrency wallet DApps.

Rug Pull scam detection involves identifying past occurrences of such events. Current research predominantly relies on machine learning methodologies for analyzing transaction records. Mazorra et al.~\cite{mazorra2022not} determine the occurrence of a Rug Pull by examining significant alterations in token price. Similarly, Xia et al.~\cite{xia2021trade} detect malicious tokens by analyzing the resemblance of token names to those traded on centralized exchanges. However, these works are limited to the retrospective detection of Rug Pull incidents and can not support preemptive analysis.

Detection of privileged nodes involves identifying high-privileged accounts that can invoke specific functions from smart contracts and transactions. This can aid auditors in providing testimony about asset ownership. Ethpector~\cite{frowis2023detecting} can uncover ownership structures from binary smart contract code on the Ethereum platform. SPCon~\cite{liu2022finding} implements role mining and security policy validation using the historical transactions of smart contracts. However, these works only focus on detecting high-privileged nodes without further examining potential centralization defects.

\section{Conclusion} \label{sec:conclusion}
This paper is structured into two main sections: the definition and detection of centralization defects. In the definition phase, we collected 597 posts from Ethereum Stack Exchange and 117 security audit reports that discuss centralization defects in smart contracts. Through manual analysis and the open card sorting method, we identified and defined six distinct types of centralization defects. Additionally, for each defect, we provide a code example along with possible solutions.

To identify centralization defects in real-world smart contracts, we developed CDRipper, a tool designed to detect the six types of defects. CDRipper constructs a permission dependency graph (PDG) and extracts the permission dependencies of functions from the source code of smart contracts.
Subsequently, it detects the sensitive operations related to centralization defects and identifies defects based on predefined patterns. We conducted a large-scale experiment employing CDRipper on 244,424 real-world smart contracts. The results were evaluated by random sampling and manual identification. Our findings indicate that 82,446 contracts exhibit at least one centralization defect, with CDRipper achieving an impressive overall precision rate of 93.7\%.
In future work, developing an automated method for generating detection rules could lower usage barriers and simplify updates for CDRipper.

\bibliographystyle{IEEEtran}
\bibliography{conference_101719.bbl}

\begin{thebibliography}{10}
\providecommand{\url}[1]{#1}
\csname url@samestyle\endcsname
\providecommand{\newblock}{\relax}
\providecommand{\bibinfo}[2]{#2}
\providecommand{\BIBentrySTDinterwordspacing}{\spaceskip=0pt\relax}
\providecommand{\BIBentryALTinterwordstretchfactor}{4}
\providecommand{\BIBentryALTinterwordspacing}{\spaceskip=\fontdimen2\font plus
\BIBentryALTinterwordstretchfactor\fontdimen3\font minus \fontdimen4\font\relax}
\providecommand{\BIBforeignlanguage}[2]{{%
\expandafter\ifx\csname l@#1\endcsname\relax
\typeout{** WARNING: IEEEtran.bst: No hyphenation pattern has been}%
\typeout{** loaded for the language `#1'. Using the pattern for}%
\typeout{** the default language instead.}%
\else
\language=\csname l@#1\endcsname
\fi
#2}}
\providecommand{\BIBdecl}{\relax}
\BIBdecl

\bibitem{certikreport2023}
\BIBentryALTinterwordspacing
``Hack3d: The web3 security quarterly report - q3 2023,'' 2023. [Online]. Available: \url{https://www.certik.com/zh-CN/resources/blog/1dloJV023Tm4ajXiluRctb-hack3d-the-web3-security-quarterly-report-q3-2023}
\BIBentrySTDinterwordspacing

\bibitem{certikblog2022}
\BIBentryALTinterwordspacing
``What is centralization risk?'' 2023. [Online]. Available: \url{https://www.certik.com/zh-CN/resources/blog/What-is-centralization-risk}
\BIBentrySTDinterwordspacing

\bibitem{mint_def}
\BIBentryALTinterwordspacing
``Minting crypto,'' 2024. [Online]. Available: \url{https://corporatefinanceinstitute.com/resources/cryptocurrency/minting-crypto/}
\BIBentrySTDinterwordspacing

\bibitem{moti_example}
\BIBentryALTinterwordspacing
``The vanishing act: How exit scammers mint new tokens undetected,'' 2024. [Online]. Available: \url{https://www.certik.com/zh-CN/resources/blog/the-vanishing-act-how-exit-scammers-mint-new-tokens-undetected}
\BIBentrySTDinterwordspacing

\bibitem{chen2020defining}
J.~Chen, X.~Xia, D.~Lo, J.~Grundy, X.~Luo, and T.~Chen, ``Defining smart contract defects on ethereum,'' \emph{IEEE Transactions on Software Engineering}, vol.~48, no.~1, pp. 327--345, 2020.

\bibitem{yang2023definition}
S.~Yang, J.~Chen, and Z.~Zheng, ``Definition and detection of defects in nft smart contracts,'' \emph{arXiv preprint arXiv:2305.15829}, 2023.

\bibitem{etherstack}
\BIBentryALTinterwordspacing
``Ethereum stack exchange,'' 2023. [Online]. Available: \url{https://ethereum.stackexchange.com/}
\BIBentrySTDinterwordspacing

\bibitem{szabo1997formalizing}
N.~Szabo, ``Formalizing and securing relationships on public networks,'' \emph{First monday}, 1997.

\bibitem{zheng2020overview}
Z.~Zheng, S.~Xie, H.-N. Dai, W.~Chen, X.~Chen, J.~Weng, and M.~Imran, ``An overview on smart contracts: Challenges, advances and platforms,'' \emph{Future Generation Computer Systems}, vol. 105, pp. 475--491, 2020.

\bibitem{evm}
\BIBentryALTinterwordspacing
``Ethereum virtual machine (evm),'' 2023. [Online]. Available: \url{https://ethereum.org/en/developers/docs/evm/}
\BIBentrySTDinterwordspacing

\bibitem{statvar}
\BIBentryALTinterwordspacing
``Solidity – state variables,'' 2023. [Online]. Available: \url{https://www.geeksforgeeks.org/solidity-state-variables/}
\BIBentrySTDinterwordspacing

\bibitem{yan2023bad}
K.~Yan, J.~Zhang, X.~Liu, W.~Diao, and S.~Guo, ``Bad apples: Understanding the centralized security risks in decentralized ecosystems,'' in \emph{Proceedings of the ACM Web Conference 2023}, 2023, pp. 2274--2283.

\bibitem{privarkey}
\BIBentryALTinterwordspacing
``Private key: What it is, how it works, best ways to store,'' 2023. [Online]. Available: \url{https://www.investopedia.com/terms/p/private-key.asp}
\BIBentrySTDinterwordspacing

\bibitem{multisig}
\BIBentryALTinterwordspacing
``Single signature, multisig or multi-party computation: How different crypto wallets protect your transactions,'' 2023. [Online]. Available: \url{https://www.ceffu.com/support/announcements/article/single-signature-multisig-multi-party-computation-how-crypto-wallets-protect-your-transactions}
\BIBentrySTDinterwordspacing

\bibitem{certik}
\BIBentryALTinterwordspacing
``Certik skynet,'' 2023. [Online]. Available: \url{https://skynet.certik.com/zh-CN}
\BIBentrySTDinterwordspacing

\bibitem{sourcehat}
\BIBentryALTinterwordspacing
``Source hat,'' 2023. [Online]. Available: \url{https://sourcehat.com/}
\BIBentrySTDinterwordspacing

\bibitem{spencer2009card}
D.~Spencer, \emph{Card sorting: Designing usable categories}.\hskip 1em plus 0.5em minus 0.4em\relax Rosenfeld Media, 2009.

\bibitem{post_oracle}
\BIBentryALTinterwordspacing
``What is the oracle problem definition exactly and briefly?'' 2023. [Online]. Available: \url{https://ethereum.stackexchange.com/questions/57071/what-is-the-oracle-problem-definition-exactly-and-briefly}
\BIBentrySTDinterwordspacing

\bibitem{report_cfc}
\BIBentryALTinterwordspacing
``Cfc project,'' 2023. [Online]. Available: \url{https://skynet.certik.com/zh-CN/projects/cfc-project}
\BIBentrySTDinterwordspacing

\bibitem{proxy}
\BIBentryALTinterwordspacing
``Proxy upgrade pattern,'' 2023. [Online]. Available: \url{https://docs.openzeppelin.com/upgrades-plugins/1.x/proxies}
\BIBentrySTDinterwordspacing

\bibitem{selfdestruct}
\BIBentryALTinterwordspacing
``Solidity selfdestruct,'' 2023. [Online]. Available: \url{https://docs.soliditylang.org/en/latest/introduction-to-smart-contracts.html\#deactivate-and-self-destruct}
\BIBentrySTDinterwordspacing

\bibitem{basicblock}
\BIBentryALTinterwordspacing
``Basic block,'' 2023. [Online]. Available: \url{https://en.wikipedia.org/wiki/Basic_block}
\BIBentrySTDinterwordspacing

\bibitem{feist2019slither}
J.~Feist, G.~Grieco, and A.~Groce, ``Slither: a static analysis framework for smart contracts,'' in \emph{2019 IEEE/ACM 2nd International Workshop on Emerging Trends in Software Engineering for Blockchain (WETSEB)}.\hskip 1em plus 0.5em minus 0.4em\relax IEEE, 2019, pp. 8--15.

\bibitem{bodell2023proxy}
W.~E. Bodell~III, S.~Meisami, and Y.~Duan, ``Proxy hunting: understanding and characterizing proxy-based upgradeable smart contracts in blockchains,'' in \emph{32nd USENIX Security Symposium (USENIX Security 23)}, 2023, pp. 1829--1846.

\bibitem{dataset}
\BIBentryALTinterwordspacing
``smart-contract-sanctuary,'' 2023. [Online]. Available: \url{https://github.com/tintinweb/smart-contract-sanctuary}
\BIBentrySTDinterwordspacing

\bibitem{sendner2024large}
C.~Sendner, L.~Petzi, J.~Stang, and A.~Dmitrienko, ``Large-scale study of vulnerability scanners for ethereum smart contracts,'' in \emph{2024 IEEE Symposium on Security and Privacy (SP)}.\hskip 1em plus 0.5em minus 0.4em\relax IEEE Computer Society, 2024, pp. 220--220.

\bibitem{confidence}
\BIBentryALTinterwordspacing
``Confidence interval - wikipedia,'' 2023. [Online]. Available: \url{https://en.wikipedia.org/wiki/Confidence\_interval}
\BIBentrySTDinterwordspacing

\bibitem{samplesizecal}
\BIBentryALTinterwordspacing
``Sample size calculator,'' 2023. [Online]. Available: \url{https://www.surveysystem.com/sscalc.html}
\BIBentrySTDinterwordspacing

\bibitem{jiang2018contractfuzzer}
B.~Jiang, Y.~Liu, and W.~K. Chan, ``Contractfuzzer: Fuzzing smart contracts for vulnerability detection,'' in \emph{Proceedings of the 33rd ACM/IEEE International Conference on Automated Software Engineering}, 2018, pp. 259--269.

\bibitem{kalra2018zeus}
S.~Kalra, S.~Goel, M.~Dhawan, and S.~Sharma, ``Zeus: analyzing safety of smart contracts.'' in \emph{Ndss}, 2018, pp. 1--12.

\bibitem{luu2016making}
L.~Luu, D.-H. Chu, H.~Olickel, P.~Saxena, and A.~Hobor, ``Making smart contracts smarter,'' in \emph{Proceedings of the 2016 ACM SIGSAC conference on computer and communications security}, 2016, pp. 254--269.

\bibitem{fp_con}
\BIBentryALTinterwordspacing
``Etherscan: Contract 0x0938095e8b4dd192756ee581e8842b7b30e69c11,'' 2023. [Online]. Available: \url{https://etherscan.io/address/0x0938095E8B4dD192756eE581e8842B7B30e69c11\#code}
\BIBentrySTDinterwordspacing

\bibitem{erc20}
\BIBentryALTinterwordspacing
``Solidity funciton modifier,'' 2023. [Online]. Available: \url{https://ethereum.org/en/developers/docs/standards/tokens/erc-20/}
\BIBentrySTDinterwordspacing

\bibitem{lamby2023trusting}
M.~Lamby, V.~Zieglmeier, and C.~Ziegler, ``Trusting a smart contract means trusting its owners: Understanding centralization risk,'' in \emph{2023 5th Conference on Blockchain Research \& Applications for Innovative Networks and Services (BRAINS)}.\hskip 1em plus 0.5em minus 0.4em\relax IEEE, 2023, pp. 1--4.

\bibitem{chen2021defectchecker}
J.~Chen, X.~Xia, D.~Lo, J.~Grundy, X.~Luo, and T.~Chen, ``Defectchecker: Automated smart contract defect detection by analyzing evm bytecode,'' \emph{IEEE Transactions on Software Engineering}, vol.~48, no.~7, pp. 2189--2207, 2021.

\bibitem{chen2017under}
T.~Chen, X.~Li, X.~Luo, and X.~Zhang, ``Under-optimized smart contracts devour your money,'' in \emph{2017 IEEE 24th international conference on software analysis, evolution and reengineering (SANER)}.\hskip 1em plus 0.5em minus 0.4em\relax IEEE, 2017, pp. 442--446.

\bibitem{tsankov2018securify}
P.~Tsankov, A.~Dan, D.~Drachsler-Cohen, A.~Gervais, F.~Buenzli, and M.~Vechev, ``Securify: Practical security analysis of smart contracts,'' in \emph{Proceedings of the 2018 ACM SIGSAC conference on computer and communications security}, 2018, pp. 67--82.

\bibitem{brent2020ethainter}
L.~Brent, N.~Grech, S.~Lagouvardos, B.~Scholz, and Y.~Smaragdakis, ``Ethainter: a smart contract security analyzer for composite vulnerabilities,'' in \emph{Proceedings of the 41st ACM SIGPLAN Conference on Programming Language Design and Implementation}, 2020, pp. 454--469.

\bibitem{nikolic2018finding}
I.~Nikoli{\'c}, A.~Kolluri, I.~Sergey, P.~Saxena, and A.~Hobor, ``Finding the greedy, prodigal, and suicidal contracts at scale,'' in \emph{Proceedings of the 34th annual computer security applications conference}, 2018, pp. 653--663.

\bibitem{frank2020ethbmc}
J.~Frank, C.~Aschermann, and T.~Holz, ``$\{$ETHBMC$\}$: A bounded model checker for smart contracts,'' in \emph{29th USENIX Security Symposium (USENIX Security 20)}, 2020, pp. 2757--2774.

\bibitem{grieco2020echidna}
G.~Grieco, W.~Song, A.~Cygan, J.~Feist, and A.~Groce, ``Echidna: effective, usable, and fast fuzzing for smart contracts,'' in \emph{Proceedings of the 29th ACM SIGSOFT International Symposium on Software Testing and Analysis}, 2020, pp. 557--560.

\bibitem{albert2018ethir}
E.~Albert, P.~Gordillo, B.~Livshits, A.~Rubio, and I.~Sergey, ``Ethir: A framework for high-level analysis of ethereum bytecode,'' in \emph{International symposium on automated technology for verification and analysis}.\hskip 1em plus 0.5em minus 0.4em\relax Springer, 2018, pp. 513--520.

\bibitem{hildenbrandt2018kevm}
E.~Hildenbrandt, M.~Saxena, N.~Rodrigues, X.~Zhu, P.~Daian, D.~Guth, B.~Moore, D.~Park, Y.~Zhang, A.~Stefanescu \emph{et~al.}, ``Kevm: A complete formal semantics of the ethereum virtual machine,'' in \emph{2018 IEEE 31st Computer Security Foundations Symposium (CSF)}.\hskip 1em plus 0.5em minus 0.4em\relax IEEE, 2018, pp. 204--217.

\bibitem{amani2018towards}
S.~Amani, M.~B{\'e}gel, M.~Bortin, and M.~Staples, ``Towards verifying ethereum smart contract bytecode in isabelle/hol,'' in \emph{Proceedings of the 7th ACM SIGPLAN international conference on certified programs and proofs}, 2018, pp. 66--77.

\bibitem{ma2023pied}
F.~Ma, M.~Ren, L.~Ouyang, Y.~Chen, J.~Zhu, T.~Chen, Y.~Zheng, X.~Dai, Y.~Jiang, and J.~Sun, ``Pied-piper: Revealing the backdoor threats in ethereum erc token contracts,'' \emph{ACM Transactions on Software Engineering and Methodology}, vol.~32, no.~3, pp. 1--24, 2023.

\bibitem{mazorra2022not}
B.~Mazorra, V.~Adan, and V.~Daza, ``Do not rug on me: Leveraging machine learning techniques for automated scam detection,'' \emph{Mathematics}, vol.~10, no.~6, p. 949, 2022.

\bibitem{xia2021trade}
P.~Xia, H.~Wang, B.~Gao, W.~Su, Z.~Yu, X.~Luo, C.~Zhang, X.~Xiao, and G.~Xu, ``Trade or trick? detecting and characterizing scam tokens on uniswap decentralized exchange,'' \emph{Proceedings of the ACM on Measurement and Analysis of Computing Systems}, vol.~5, no.~3, pp. 1--26, 2021.

\bibitem{frowis2023detecting}
M.~Fr{\"o}wis and R.~B{\"o}hme, ``Detecting privileged parties on ethereum,'' in \emph{International Conference on Financial Cryptography and Data Security}.\hskip 1em plus 0.5em minus 0.4em\relax Springer, 2023, pp. 470--488.

\bibitem{liu2022finding}
Y.~Liu, Y.~Li, S.-W. Lin, and C.~Artho, ``Finding permission bugs in smart contracts with role mining,'' in \emph{Proceedings of the 31st ACM SIGSOFT International Symposium on Software Testing and Analysis}, 2022, pp. 716--727.

\end{thebibliography}


\end{document}